\documentclass[a4paper,showpacs,nofootinbib,twocolumn]{revtex4}
\usepackage[utf8]{inputenc}
\usepackage{graphicx,bm,color,amssymb,amsmath,natbib,epsfig,hhline}
\usepackage{graphicx}
\usepackage{slashed,verbatim,ulem}
\definecolor{mypink}{RGB}{219, 48, 222}

\usepackage{subfigure}
\usepackage[colorlinks=true,linktocpage=true,linkcolor=blue,citecolor=blue]{hype
rref}

\def\be {\begin{equation}}
\def\ee {\end{equation}}
\def\bea {\begin{eqnarray}}
\def\eea {\end{eqnarray}}
\def\bc {\begin{center}}
\def\ec {\end{center}}
\def\nn {\nonumber}

\def\zbf#1{{\bf {#1}}}
\def\bfm#1{\mbox{\boldmath $#1$}}
\date{\today}

\begin{document}
\title{Impact of Magnetic field on neutron star properties}

\author{Tuhin Malik$^1$}
\email{tuhin.malik@gmail.com}
\author{Debashree Sen$^1$}
\email{debashreesen88@gmail.com }
\author{T. K. Jha$^1$}
\email{tkjha@goa.bits-pilani.ac.in}
\author{Hiranmaya Mishra $^2$}
\email{hm@prl.res.in }

\affiliation{ $^1$Department of Physics, BITS Pilani, K.K. Birla Goa Campus, GOA - 403726, India. \\
$^2$Theoretical Physics Division, Physical Research Laboratory, Navrangpura, Ahmedabad-380 009, India}

\begin{abstract}
We derive an equation of state for magnetized charge neutral nuclear matter relevant for neutron star 
structure. The calculations are performed within an effective chiral model based on generalization of 
sigma model with nonlinear self interactions of the sigma mesons along with vector mesons and a 
$\rho-\sigma$ cross-coupling term. The effective chiral model is extended by introducing the 
contributions of strong magnetic field on the charged particles of the model. The contributions arising 
from the effects of magnetic field on the Dirac sea of charged baryons are also included. The resulting 
equation of state for the  magnetized dense matter is used to investigate the neutron star properties, 
like, mass-radius relation and tidal deformability. The dimensionless tidal deformability of 
$1.4~{M}_\odot$ NS is found to be $\Lambda_{1.4}=526$, which is consistent with recent observation 
of GW170817. The maximum mass of neutron star in presence of strong magnetic field is consistent with 
the observational constraints on mass of neutron star from PSR~ J0348 - 0432 and the radius at 
$1.4~{M}_\odot$ mass of the neutron star is within the empirical bounds. 
\end{abstract}

\maketitle
\newpage

                    
\section{Introduction}  
         The extreme properties of neutron stars (NSs) not only opens up many possibilities 
         related to the composition, structure and dynamics of stable cold matter in the observable 
         universe but also to the matter interaction at the fundamental level. Almost every aspect 
         of them, be it mass, radius, rotational frequency or the magnetic field, represents matter 
         at extreme conditions. The structure of NS depends on nuclear equation of state (EOS), 
         which is poorly known till date. Although the observation of high mass pulsars like PSR ~J1614-2230 
         ($M = 1.928 \pm~ 0.04 M_{\odot}$) \cite{Fonseca2016} and PSR~ J0348 - 0432 ($M = 2.01 \pm~ 0.04~ M_{\odot}$)  
         \cite{Antoniadis2013} severely constrains the EOS and their interaction as well but also 
         questions the possible presence of exotic matter in them. Apart from the observational 
         constrains from high mass stars, the  event of binary neutron star merger observed gravitational 
         waves GW170817 in August 2017 disfavors some of the stiff EOSs \cite{GW170817}. The precise 
         knowledge of NS radius can also constrains the behavior of EOSs. The empirical estimates of radius 
         of a canonical NS ($M = 1.4 M_{\odot}$) should be $R_{1.4} = (11.9 \pm 1.22)~ $ km \cite{Lattimer2013}. 
         Recently in Refs. \cite{Fattoyev2018,Malik2018}, using the extracted bounds on neutron star tidal 
         deformability of GW170817 event suggests that $R_{1.4} < 13.76$ km.  

         Born out of massive interstellar gases over millions of years, the magnetic field present in 
         these compact structures can be very high, although the origin of these high fields is still 
         not well understood. The typical values of surface magnetic field of neutron stars ranges from $10^{12}$ 
         to $10^{15}$ Gauss. It is speculated that the field intensity can be even more at the core. 
         A fraction of the population, having the strongest surface magnetic fields $\sim$ ($10^{12} - 10^{15})$ 
         Gauss are called the magnetars and generally they belong to Soft Gamma Repeaters (SGRs) or 
         Anomalous X-ray Pulsars (AXPs) \cite{Gomes2017}. Typical examples of such magnetars are 
         the $1{\rm E}~1048.1-5937$ and $1{\rm E}~2259+586$ with surface magnetic field $B_{{\rm surf}}\sim 10^{14}$ 
         Gauss \cite{Melatos1999}, $4{\rm U}~0142+61$ ($B_{{\rm surf}} \sim 10^{16}$ Gauss) 
         \cite{Makishima2014} and ${\rm SGR}~1806-20$ ($B_{{ \rm surf} }\sim 10^{14}$ Gauss) \cite{Kouveliotou1998} etc. 
         One can refer to the magnetar catalog available online \cite{Olausen2014} for more such examples. 
         It is conceivable, though perhaps speculative, that in the interior of the magnetars, the magnetic 
         field could be several orders of magnitude larger and therefore is expected to affect the dense 
         matter properties on the scale of QCD. \cite{Lai1991} 
         Therefore, it is desirable to  incorporate the magnetic field effects to determine the composition 
         and the gross structural properties of these neutron stars with high magnetic field. Various authors 
         have incorporated the effects of magnetic field in neutron stars 
         \cite{Chakrabarty1997,Broderick2002,Huang2010,Lopes2012,Casali2014,Gomes2014,Gomes2017,
         Gao2015,Orsaria2011,Dexheimer2014,Franzon2015,Chu2015,Paulucci2011,Strickland2012} to account for 
         the properties of neutron stars. From various investigations, it is well known that the strong 
         magnetic fields affects the energy levels of charged particles due to Landau quantization, 
         which may be strong enough to make the pressure of the matter anisotropic and therefore 
         spherical symmetry may not be a suitable approximation in order to study the structural 
         properties of neutron stars \cite{PerezMartinez2008,Strickland2012}. However, it has been noted 
         that the difference in NS properties (such as mass and radius), calculated separately for parallel 
         and perpendicular directions to the magnetic field being small, spherical symmetry can still hold 
         \cite{Chu2015,Huang2010}. In most of these calculations, the divergent vacuum contribution is omitted. 
         Indeed, for vanishing magnetic fields such a 'no sea approximation', leads to a very small 
         difference in the EOS as compared to the EOS calculated taking into account the Dirac sea effect 
         after re-normalization \cite{Glendenning1988,Glendenning1989}.
         We shall include here the effects of the magnetic field on the Dirac sea of nucleons. 
         In fact, inclusion of magnetic field effects for the Dirac vacuum has been the reason for magnetic 
         catalysis of chiral symmetry breaking in quark matter and has been studied in various effective models 
         like Nambu-Jona-Lasinio (NJL) models\cite{Menezes2009,Chatterjee2011}  
         as well as quark-meson models\cite{Ferrari2012,Andersen2012,Skokov2012}. 
         Apart from the static structural properties of the magnetars, magnetic fields also play an important 
         role in the physics 
         of compact star mergers. The gravitational waves emitted at the late stage of the merger process 
         can possibly be detected directly and are sensitive to the EOS of the dense matter \cite{Read2013}. 
         The magnetic field in a merger process can possibly become extremely large due to magneto-rotational 
         instability and the magnitude can be large enough to affect the EOS of dense matter.

          With this motivation, in the present work, we incorporate the effects of strong magnetic 
          field on the EOS and calculate the NS properties using a model based on a generalization 
          of sigma model. In such a model, the nucleons are coupled with the $\sigma$ and pion fields 
          along with a potential for the $\sigma,{\bfm\pi}$ fields. Such a model, generalized to include 
          vector mesons coupled to the scalar fields to give masses to vector mesons, was used to study 
          finite temperature aspects of nuclear matter and its applications to NS \cite{Glendenning1986}. 
          Later on, it was generalized to include iso-vector rho meson as well as higher order 
          non-linear meson interactions \cite{Sahu2000,Jha2006}. The reason was to 
          include the effects of isospin asymmetry as well as to explain the rather high value of 
          the nuclear incompressibility that one gets when non-linear meson interactions are not taken 
          into account. Further, the model has been generalized recently \cite{Malik2017}, to include 
          cross coupling between isovector and isoscalar mesons. These cross couplings were found to be 
          instrumental in explaining the density dependence of the nuclear symmetry energy as well as its 
          slope and curvature parameters at the saturation density deduced from diverse set of 
          experimental data. The EOS was also used to explore the gross structural properties of NS like 
          mass and radius which turned out to be consistent with measurement for the maximum mass while the 
          radius at the canonical mass is within the empirical bounds \cite{Malik2017}. 
          The present investigation of the effects of  strong magnetic field is carried 
          out within this effective chiral model.

          The paper is organized as follows. In the next section, we first recapitulate the essential 
          features of the effective chiral model. In the next subsection we introduce the magnetic 
          field in the model and calculate the pressure and energy density in presence of external magnetic 
          field in some  detail. As we shall see, in presence of a strong magnetic field, the EOS can 
          become anisotropic when the induced magnetization effects become strong. After deriving the EOS, 
          in section III, we solve the Tolman-Oppenheimmer-Volkoff (TOV) equations using the EOS so 
          derived in presence of magnetic field. We also consider here the tidal deformation of NS in 
          the context of NS mergers. The results along with brief discussions are presented in section IV. 
          Finally, we  summarize and present an out look of the present investigation in section V.

\section{The effective chiral model and equation of state}
                  We now discuss briefly the salient features of the effective chiral Lagrangian of 
                  Ref.\cite{Malik2017} to describe dense nuclear matter. The effective Lagrangian of 
                  the model interacting through the exchange of the pseudo-scalar meson $\pi$, 
                  the scalar meson $\sigma$, the vector meson $\omega$ and the iso-vector $\rho-$meson 
                  is given by
\begin{widetext}
\begin{eqnarray}
\label{eq1}
{\cal L}&=& \bar\psi_{B}~\left[ \big(i\gamma_\mu\partial^\mu
         - g_{\omega}\gamma_\mu\omega^\mu
         - \frac{1}{2}g_{\rho}{\bfm \rho}_\mu\cdot{\bfm \tau}
            \gamma^\mu\big )
         - g_{\sigma~}~\big(\sigma + i\gamma_5
             \bfm \tau\cdot\bfm \pi \big)\right]~ \psi_{B}
\nonumber \\
&&
        + \frac{1}{2}\big(\partial_\mu\bfm \pi\cdot\partial^\mu\bfm\pi
        + \partial_{\mu} \sigma \partial^{\mu} \sigma\big)
        - \frac{\lambda}{4}\big(x^2 - x^2_0\big)^2
        - \frac{\lambda b}{6 m^2}\big(x^2 - x^2_0\big)^3
        - \frac{\lambda c}{8 m^4}\big(x^2 - x^2_0\big)^4
\nonumber \\
&&      - \frac{1}{4} F_{\mu\nu} F_{\mu\nu}
        + \frac{1}{2}{g_{\omega B}}^{2}x^2 \omega_{\mu}\omega^{\mu}
        - \frac {1}{4}{\zbf R}_{\mu\nu}\cdot{\zbf R}^{\mu\nu}
        + \frac{1}{2}m^{\prime 2}_{\rho}{\bfm \rho}_{\mu}\cdot{\bfm \rho}^{\mu}\nonumber\\
&& +\eta_1\left(\frac{1}{2}g_\rho^2x^2\bfm\rho_\mu\cdot\bfm\rho^\mu\right)+\eta_2\left(\frac{1}{2}g_\rho^2x^2\bfm\rho_\mu
\cdot\bfm\rho^\mu\omega_\mu\omega^\mu\right).
\end{eqnarray}
\end{widetext}

      The first line of the above Lagrangian represents the interaction of the nucleon isospin doublet 
      $\psi_B$ with the  mesons. In the second line we have the kinetic and the non-linear terms in 
      the pseudo-scalar-isovector pion field '$\bfm \pi$', the scalar field '$\sigma$', and higher 
      order terms of the scalar field in terms of the chiral 'invariant combination of the two i.e., 
      $x^2= {\bfm \pi}^2+\sigma^{2}$. In the third line, we have the field strength and the mass term 
      for the vector field '$\omega$' and the iso-vector field '$\bfm \rho$' meson. The last line contains 
      the cross coupling terms between $\bfm \rho$ and $\omega$ and also between $\bfm\rho$ and $\sigma$ mesons. 
      $g_{\sigma}, g_{\omega}$ and $g_{\rho}$ are the usual meson-nucleon coupling strength of the scalar, 
      vector and the iso-vector fields respectively. Here we shall be concerned only with the normal non-pion 
      condensed state of matter, so we take $<\bfm \pi>=0$ and also $m_{\pi} = 0$. The last two terms in
      the Lagrangian incorporates the effect of cross-couplings between $\rho-\sigma$ and $\rho-\omega$ with 
      coupling strengths $\eta_1~g_{\rho}^2$ and $\eta_2~g_{\rho}^2$ respectively. From our previous work 
      \cite{Malik2017}, where we investigated the role of cross-coupling terms to constrain symmetry 
      energy and NS properties, we concluded that the inclusion of $\rho-\sigma$ cross-coupling term 
      is sufficient to satisfy the overall properties. Hence in the present work, we only consider 
      the $\rho-\sigma$ coupling \cite{Malik2017} with coupling parameter $\eta_1$.

      The interaction of the scalar and the pseudoscalar mesons with the vector boson generates a dynamical 
      mass for the vector bosons through spontaneous breaking of the chiral symmetry with scalar field attaining 
      the vacuum expectation value $x_0$. Then the mass of the nucleon ($m$), the scalar ($m_{\sigma}$) 
      and the vector meson mass ($m_{\omega}$), are related to $x_0$ through
\begin{eqnarray}
m = g_{\sigma} x_0,~~ m_{\sigma} = \sqrt{2\lambda} x_0,~~
m_{\omega} = g_{\omega} x_0\ .
\end{eqnarray}
\noindent
where, $\lambda=\frac{(m_\sigma^2-m\pi^2)}{2f_\pi^2}$ and $f_\pi=x_0$ is the pion decay constant 
reflecting the strength of SSB. Due to the cross coupling between $\bfm\rho$ and $\sigma$ mesons, 
there is a contribution to the $\bfm\rho$-meson mass from the vacuum expectation value of 
the $\sigma$ meson: i.e. $m_\rho^2=m^{\prime 2}_\rho+ \eta_1g_\rho^2 x_0^2$.

       To obtain the EOS, we revert to the mean-field procedure, where, one assumes the mesonic fields 
       to be classical and uniform mean fields  while retaining the quantum nature of the baryonic field 
       i.e.$\langle\sigma\rangle=\sigma$, $\langle\omega_\mu\rangle=\omega_0\delta_{\mu 0}$, 
       $\langle\rho_\mu^a\rangle$ =$\delta_{\mu 0}\delta_{a 3}\rho_{03}$.

       We recall here that this approach has been extensively used to obtain field-theoretical EOS for 
       high density matter \cite{Sahu2000,Sahu2004,Jha2009},
       and gets increasingly valid when the source terms are large \cite{Serot1986}. 
       The details of the present model and its attributes such as the derivation of the equation of 
       motion of the meson fields and its equation of state $(\varepsilon~\&~P)$ can be found in  
       Ref. \cite{Malik2017}. For the sake of completeness however, we write down the meson field 
       equations in the mean-field ansatz. Moreover these mean field equations remain the same
       even in presence of magnetic field which we discuss in the next subsection.
       The mean meson fields i.e. the vector field ($\omega$), 
       scalar field ($\sigma$) and isovector field ($\rho_3^0$) are determined by solving the mean 
       field equations which (in terms of $Y = x/x_0 = m^*/m$) are, respectively, given by
\begin{eqnarray}
\label{eq3}
&& \Big[m_\omega^2 Y^2 + \eta_2 C_\rho m_\rho^2 (\rho_3^0)^2\Big] \omega_0 = g_\omega \rho, \\   
\label{eq4}
&&(1-Y^2)-\frac{b}{m^2 C_\omega}(1-Y^2)^2+\frac{c}{m^4 C_{\omega}^2}(1-Y^2)^3 \nonumber \\
&&+\frac{2 C_\sigma m_\omega^2 \omega_0^2}{m^2}+\frac{2 \eta_1 C_\sigma C_\rho m_\rho^2 (\rho_3^0)^2}{C_\omega m^2}
     - \frac{2 C_\sigma \rho_s}{m Y}=0 ,\\
\label{eq5}
&& m_\rho^2\Big[1 - \eta_1 (1-Y^2)C_\rho/C_\omega + \eta_2 C_\rho \omega_0^2\Big]\rho_3^0 \nonumber \\
&&~~~~= \frac{1}{2} g_\rho (\rho_p - \rho_n).
\end{eqnarray}
The quantity $\rho$ and $\rho_S$ are the baryon and the scalar density defined as,
\begin{eqnarray}
\label{density}
    &&\rho = \frac{\gamma}{(2 \pi)^3} \sum_{i=n,p} \int_0^{k_F^i}d\zbf k,  \\
    &&\rho_s = \frac{\gamma}{(2 \pi)^3} \sum_{i=n,p} \int_0^{k_F^i}\frac{m^*}{\sqrt{m^*{^2}+ \zbf k^2}} d\zbf k,    
\end{eqnarray}
where, $k_F^i$ is the Fermi momentum of the nucleon and $\gamma=2$ 
is the spin degeneracy factor. $m^*=g_\sigma x$ is the medium dependent 
nucleon mass.  $C_\sigma\equiv g_\sigma^2/m_\sigma^2$ , $C_\omega\equiv g_\omega^2/m_\omega^2$ 
and $C_\rho\equiv g_\rho^2/m_\rho^2$ are the scalar, vector and isovector coupling parameters 
which enter into the actual computation. These parameters are given in Table \ref{tab1}.  

The energy density and the pressure of the considered model is given by,
{\small 
\begin{eqnarray}
\label{eq8}
      \epsilon &=& \frac{1}{\pi^2}\sum_{i=n,p}\int_0^{k_F^i}k^2\sqrt{k^2+m^*{^2}}dk 
                       + \frac{m^2}{8 C_{\sigma}}(1-Y^2)^2 \nonumber\\
      &&- \frac{b}{12 C_{\sigma}C_{\omega}}(1-Y^2)^3 + \frac{c}{16 m^2 C_{\sigma}
            C_{\omega}^2}(1-Y^2)^4  
        + \frac{1}{2}m_{\omega}^2\omega_{0}^2Y^2  \nonumber \\
      &&+ \frac{1}{2}m_\rho^2\Big[1- \eta_1(1-Y^2) (C_\rho/C_\omega)+ 3 \eta_2 C_\rho 
           \omega_0^2\Big] (\rho_3^0)^2, \\
\label{eq9}
   p &=& \frac{1}{3 \pi^2}\sum_{i=n,p}\int_0^{k_F^i}\frac{k^4}{\sqrt{k^2+m^*{^2}}}dk 
         - \frac{m^2}{8 C_{\sigma}}(1-Y^2)^2 \nonumber \\
     &&+ \frac{b}{12 C_{\sigma}C_{\omega}}(1-Y^2)^3 - \frac{c}{16 m^2 
            C_{\sigma}C_{\omega}^2}(1-Y^2)^4+ \frac{1}{2}m_{\omega}^2\omega_{0}^2Y^2     \nonumber\\ 
     && + \frac{1}{2}m_\rho^2 \Big[1- \eta_1(1-Y^2) (C_\rho/C_\omega)+ \eta_2 C_\rho 
              \omega_0^2\Big] (\rho_3^0)^2 .
\end{eqnarray}}
Clearly in the above, we have neglected the contributions of Dirac sea of nucleons and have 
kept the contribution arising from the Fermi sea of nucleons with an effective mass $m^*$ given 
by the first terms in Eqs.(\ref{eq8},\ref{eq9}).

\begin{table}
\caption{\label{tab1}Parameters of the present model such as the couplings 
$C_\sigma\equiv g_\sigma^2/m_\sigma^2$ , $C_\omega\equiv g_\omega^2/m_\omega^2$ 
and $C_\rho\equiv g_\rho^2/m_\rho^2$. $B=b/m^2$ and $C=c/m^4$ are the higher order 
scalar couplings and $\eta_1$ is the coupling strength for the $\rho-\sigma$ 
term. Below are the nuclear matter saturation properties such as nucleon 
effective mass ($m^*$), nuclear matter incompressibility ($K$), energy per 
particle ($e_0$), symmetry energy $J_0$ and symmetry energy slope parameter ($L_0$), 
all defined at nuclear matter saturation density $\rho_0=0.153~fm^{-3}$, 
taken from \cite{Malik2017}}.  
\setlength{\tabcolsep}{10pt}
\renewcommand{\arraystretch}{1.1}
\begin{tabular}{cccccc}  
\toprule
$C_\sigma$ & $C_\omega$ & $C_\rho$ & $\eta_1$ &  $B$ & $C$   \\ 
$\text{fm}^2$ & $\text{fm}^2$ & $\text{fm}^2$ &  & $\text{fm}^2$ & $\text{fm}^4$ \\
\hline
7.057 & 1.757 & 12.28 & -0.79 & -5.796 & 0.001  \\
\toprule
$\rho_0$ & $m^{\star}$ & $K$ & $e_0$ & $J_0$  & $L_0$ \\
\hline 
0.153 &  0.86  & 247  & -16.0 & 32.5  & 65  \\
\toprule
\end{tabular}
\end{table}
   Such a 'no-sea approximation' is a reasonable approximation regarding equation of 
state \cite{Glendenning1988,Glendenning1989}. However, in presence of magnetic field, the contribution 
of Dirac sea can become significant as will be discussed in the next subsection.

\subsubsection{The EOS with magnetic field}
        We shall consider here the effect of magnetic field on the equation of state as given 
        in the previous subsection. We shall consider the magnetic field to be constant and to 
        be in the z-direction without loss of generality. Further, we choose here the gauge 
        $A_\mu=\delta_{\mu 2}x B$ where $B$ is the magnitude of the magnetic field. In presence 
        of the magnetic field, the nucleon as well as the rho meson kinetic terms will get 
        modified with the derivative of the fields getting replaced by covariant derivatives. 
        In the mean field approximation, the contribution of the nucleons to the thermodynamic 
        potential $\Omega_N$ will depend upon the mass of the nucleons and the external 
        thermodynamic parameters like baryon chemical potential ($\mu$), magnetic field $B$ 
        and temperature ($T$). Since the baryon mass will be determined dynamically by 
        minimization of the thermodynamic potential, it will be dependent on these parameters 
        implicitly. We can write the thermodynamic potential 
        $\Omega_N=\Omega_N(m^*(\mu,B,T),\mu,B,T)$ (negative of the pressure) as
\be
\Omega_N=\Omega_{N_{sea}}+\Omega_{N_ {med}}
\ee
where, $\Omega_{N_{sea}}(m^*(\mu,B,T),0,B,0)$ is the free energy of the magnetized from the 
Dirac sea while $\Omega_{N,med}$ is the contribution from the Fermi sea. Let us consider 
charged nucleons i.e. protons first. With Landau quantization for the charged nucleons, 
$\Omega_{N_{sea}}$ is given explicitly as
\bea
&&\Omega_{N_{sea}}(m^*(\mu,B,T),B,\mu=0,T=0)\nonumber\\
&&  =-\frac{|qB|}{(2\pi)^2}\sum_{n=0}^{\infty}\alpha_n\int dp_z\epsilon_n(p_z)
\label{omegasea}
\eea
where, {$\epsilon_n(p_z)=\sqrt{p_z^2+2n|qB|+m^{\star 2}(B,\mu,T)}$} is the energy of the  
nucleon with charge $q$ for the n-th Landau level and $\alpha_n=2-\delta_{n0}$ is the degeneracy 
of the Landau level i.e. all levels except the lowest landau level is doubly degenerate. 
Let us note that $\Omega_{N_{sea}}(m^*(B,\mu),B,\mu=0,T=0)$ is not a vacuum term in the strict 
sense as the nucleon mass still depends on the medium. 

         The medium contribution to the thermodynamic potential at a given temperature $\beta^{-1}$ is 
given by
\bea
\Omega_{N_{med}}&=&-\sum_n\frac{|qB|}{(2\pi)^2\beta}\int dp_z
\Bigg[\log(1+e^{-\beta(\epsilon_n-\mu^\star)})\nonumber\\
&+&\log(1+e^{-\beta(\epsilon_n+\mu^\star)})\Bigg],
\label{medn}
\eea
\noindent 
where, $\mu^\star$ is the effective chemical potential of the baryon in presence of 
vector mean fields  and is given by $\mu^\star=\mu-g_\omega\omega_0-g_\rho I_3\rho_0$.
In the zero temperature limit of the $\Omega_{med}$, the anti baryonic contribution will 
vanish  and only the particle part will contribute. Using the relation $\lim_{\beta\to \infty}
(1/\beta)\log (1+e^{-\beta x})=-x\theta(-x)$, 
the integrand of Eq.(\ref{medn}) becomes $(\epsilon_n-\mu^\star)
\theta(\mu^*-\epsilon_n)$. The theta function restricts the integration 
over the variable $p_z$  up to a maximum of $p_{F}^n=\sqrt{\mu^{\star 2}-m^{\star 2}- 2n|q|B}$ for a 
given value of landau level $n$. Further, the positive value  of $p_z^2$  restricts the sum over the Landau levels 
up to a maximum $n_{max}=Int[\frac{\sqrt{\mu^{\star 2}-m^{\star 2}}}{2|q|B}]$. After the integration over 
$p_z$ ,the medium contribution is now given by
\bea
\label{eq13}
&&\Omega_{med}=\sum_{n=0}^{n_{max}} \frac{{\alpha}_n |q| B}{4\pi^2} 
\Bigg[ \mu^\star p_{F}^n\nonumber\\
& - &\Big(m^{\star 2} 
+ 2n|q| B \Big) \log \left(\frac{p_{F}^n + \mu^{\star}}{\sqrt{m^{\star 2} + 2n|q_i| B}}\right) \Bigg].
\eea

Now, let us discuss the sea contribution to the free energy given in Eq.(\ref{omegasea}). 
This integral is divergent. We regularize this with dimensional regularization. This has been 
used earlier in the context of chiral symmetry breaking in presence of magnetic field 
\cite{Menezes2009,Chatterjee2011} as well as in hadron resonance gas model \cite{Endrodi2013}. Another 
alternate method often used to regularize such divergent integrals is thorough proper time 
method yielding similar results \cite{Gusynin1994,Gusynin1995a,Gusynin1995b,Haber2014}.
To regularize $\Omega_{N_{sea}}$ one adds and subtract a zero magnetic field sea contribution\cite{Menezes2009}.
The divergent zero magnetic field part is evaluated in d=$3-\epsilon'$, while the integral over $dp_z$, in
the presence of magnetic field is evaluated in d=$1-\epsilon'$ with $\epsilon'\rightarrow 0$.
Such a manipulation results in  
\bea
&&\Omega_{N_{sea}}(B) -\Omega_{N_{sea}}(B=0)
=-\frac{|qB|^2}{2\pi^2}\bigg[\zeta'(-1,x)\nonumber\\
&&-\frac{1}{2}(x^2-x)\log(x)+\frac{x^2}{4}+\frac{1}{12}\log x\nonumber\\
&&+\frac{1}{12}
\left(-\frac{2}{\epsilon'}+\gamma_E-1+\log\frac{4\pi\mu^2}{M^2}\right)\bigg],
\label{omegasea}
\eea
where, $\gamma_E\simeq 0.577$ is the Euler-Mascheroni constant,
$\mu$ is scale related to dimensional regularization and $\zeta'(-1,x)$ is the
derivative of the the Riemann-Hurwitz $\zeta$- function $\zeta(z,x)$ at $z=-1$ and 
is given by\cite{Elizalde1985}
\bea
&&\zeta'(-1,x)=-\frac{1}{2}x\log x-\frac{x^2}{4}+\frac{1}{2}x^2\log x\nonumber\\
&&+
x^2\int_0^{\infty}\frac{2\tan^{-1}y +y\log(1+y^2)}{\exp(2\pi xy)-1} dy.
\label{zetader}
\eea
We have abbreviated here $x\equiv \frac{m^{\star ^2}}{2|qB|}$. Further, $\Omega_{N_{sea}}(B=0)$ is the
the (divergent) Dirac sea contribution to the thermodynamic potential at vanishing magnetic field.
\be
\Omega_{N_{sea}}(B=0)=\frac{\gamma}{(2\pi)^3}\int d\zbf k\sqrt{\zbf k^2+m^{\star 2}}
\ee
As noted earlier, since the zero field vacuum contribution is known to have small effects on the equation of state, 
we shall consider here solely the B-dependent sea contribution. This contribution given in Eq.(\ref{omegasea}) is still divergent, 
as it has a purely magnetic field dependent term $\sim B^2/\epsilon'$. As explicitly shown in Ref.\cite{Endrodi2013}, 
such a divergence is taken care of by  adding the pure field contribution 
$B^2/2$ to the  field dependent contribution of the Dirac vacuum. The contribution
is rendered finite by   defining renormalized  charge $q_r$ and renormalized magnetic field $B_r$ through\cite{Endrodi2013,Haber2014} 
\be 
B^2=Z_qB_r^2,\quad \quad q_r^2=Z_q^{-1} q_r^2 ,\quad\quad q_rB_r=qB
\ee
Once this is done, the free energy still depends upon the scale $\mu$ of renormalization. 
However,as explicitly shown in Ref.\cite{Haber2014}, one can choose the renormalisation scale such that the thermodynamic potential can be written only in terms of renormalized 
quantities so that
\bea
&&\Omega_{N_{sea}}+\frac{1}{2}B^2\nonumber\\
&&=\frac{B_r^2}{2} +
\frac{|q_rB_r|}{(2\pi)^2}\bigg[\zeta'(-1,x)\nonumber\\
&&-\frac{1}{2}(x^2-x)\log(x)+\frac{x^2}{4}+\frac{1}{12}\log x\bigg]\nonumber\\
&&\equiv \frac{B_r^2}{2}+\Omega_{f}.
\label{omgf}
\eea
In what follows we shall suppress the subscript $'r'$ from the magnetic field and the charge 
but it is understood that the field and charges used are the renormailzed quantities.

The Dirac vacuum contribution also affects the scalar condensate $\rho_s$ of Eq.(\ref{eq19}). 
The contribution to the scalar density from charged baryons of a given specie $'i'$, $\rho_s^i$ is given by
\bea
\label{eq19}
\rho_s^i&=&=\langle\bar\psi_i\psi_i\rangle\nonumber\\
&=&\sum_n\alpha_n\frac{|q_i|B}{(2\pi)^2}\int dp_z\frac {m_i^\star}{\epsilon_n}
\theta(\mu^\star-\epsilon_n)\nonumber\\
&-&\frac{m_i^\star |q_i|B}{2\pi^2}\left[x_i(1-\log x_i)+\log \Gamma(x_i)+\frac{1}{2}
\log(\frac{x_i}{2\pi})\right]\nonumber\\
&\equiv& \rho_s^{med}+\rho_s^{\rm field}
\eea
The theta function restricts the integration over the variable $p_z$  up to a maximum of 
$p_{F,n}^i=\sqrt{\mu^{\star 2}-m_i^{\star 2}- 2n|q_i|B}$ for a given value of $n$. Positivity of 
$p_z^2$ again restricts the sum over the Landau levels up to a maximum $n_{max}=Int[\frac{\sqrt{\mu^{\star 2}-m^{\star 2}}}{2|q_iB}]$. 
One can perform the integration of $p_z$ analytically to obtain
\be
\rho_s^{med}=\sum_{n=0}^{n_{max}}\frac{|q_i|B}{2\pi^2}\alpha_n
 m_i^{\star}\log\left(\frac{{ p_{F,n}^i+\mu^{\star}}}{\sqrt{m_i^{\star 2}+2n|q_i|B}}\right)
\ee

The number density of charged baryon of a given species similarly is given by
\be
\rho_i=\sum_n\frac{2|q_i|B}{4\pi^2}\sqrt{\mu^{\star 2}-m_i^{\star 2}-2n|q_i|B}
\ee

The meson field equations, in presence of magnetic field are same as 
given in Eqs.(\ref{eq3} ,\ref{eq4}, \ref{eq5}) except that the scalar density is now given as
\be
\rho_s=\rho_s^n+\rho_s^p
\ee
with the neutron contribution to the scalar density being
\be
\rho_s^n=\frac{m^\star}{2\pi^2}\left[k_F^n\mu_n^\star-m^{\star 2}\log\frac{\mu_n^\star+k_F^n}{m^\star}\right]
\ee
while, the proton contribution to the scalar density is given by, with $x_p=\frac{{m^*}^2}{2|eB|}$
\bea
\label{eq24}
&&\rho_s^p=
\sum_{n=0}^{n_{max}}\frac{|eB|}{2\pi^2}\alpha_n
 m^*\log\left(\frac{{ k_{F,n}^p+\mu_p^\star}}{\sqrt{m^{\star 2}+2n|eB|}}\right)\nonumber\\
&&
-\frac{m^\star |e|B}{2\pi^2}\left[x_p(1-\log x_p)+\log \Gamma(x_p)+\frac{1}{2}
\log(\frac{x_p}{2\pi})\right]\nonumber\\
\eea
Similarly, the the baryon number densities of neutron and protons are given as
\be
\rho_n=\frac{{k_{F}^n}^{3}}{3\pi^2}\quad\quad\rho_p=\sum_{n=0}^{n_{max}}\frac{|eB|}{2\pi^2}k_{F,n}^p
\label{rhopn}
\ee
where, $k_{F,n}^p=\sqrt{\mu_p^{\star 2}-m^{\star 2}- 2n|e|B}$ for a given value of the Landau level $n$.

      Next we write down the equation of state i.e. energy density and pressure in the present 
model in presence of external magnetic field. The energy density is given by
\be
\epsilon=\epsilon_n+\epsilon_p+\epsilon_{meson}+\frac{1}{2}B^2.
\label{eosenmag}
\ee
In the above, the energy density of neutrons $\epsilon_n$ is given as
\be
\epsilon_n=\frac{1}{8\pi^2}\left[k_{F}^n\mu_n^\star(2m^{\star 2}+{k_F^n}^2)-m^{\star 4}\log\left(\frac{\mu_n^\star
+k_F^n}{m^\star}\right)\right].
\ee
The contribution of the protons, on the other hand, arise from the Dirac sea as in 
Eq.(\ref{omgf}) as well as the medium,the Fermi sea of protons.
\bea
&&\epsilon_p=
-\frac{|eB|^2}{2\pi^2}\bigg[\zeta'(-1,x)\nonumber\\
&&-\frac{1}{2}(x_p^2-x_p)\log(x_p)+\frac{x_p^2}{4}+\frac{1}{12}\log x_p\bigg]\nonumber\\
&&+\frac{|eB|}{4\pi^2}\sum_n^{n_{max}}[\alpha_n (k_{F,p}^n+\mu_p^\star)\nonumber\\
&&+({m_n^p})^2\log\left(\frac{k_{F, n}^p+\mu_p^\star}{m_n^p}\right)],
\eea
where we have introduced , mass in the n-th Landau label for proton as $m_n^p=\sqrt{m^{\star 2}+2 n |eB|}$. 
The contribution to the energy 
density from the mesons arises from the potential terms of the mesons and is given by
\bea
&&\epsilon_{meson}=\frac{m^{\star 2}}{8 C_{\sigma}}(1-Y^2)^2 
     - \frac{b}{12 C_{\sigma}C_{\omega}}(1-Y^2)^3\nonumber\\
&& + \frac{c}{16 m^2 C_{\sigma}
            C_{\omega}^2}(1-Y^2)^4  
        + \frac{1}{2}m_{\omega}^2\omega_{0}^2Y^2  \nonumber \\
      &&+ \frac{1}{2}m_\rho^2\Big[1- \eta_1(1-Y^2) (C_\rho/C_\omega)\nonumber\\ &&+ 3 \eta_2 C_\rho 
           \omega_0^2\Big] (\rho_3^0)^2.
\eea

Similarly, the pressure , the negative of the thermodynamic potential can be written as
\be
P=P_n+P_p-\epsilon_{meson}-\frac{1}{2}B^2
\equiv P_0-\frac{1}{2}B^2
\label{eosprmag}
\ee
The contribution of the neutrons to the pressure, using Eq.(\ref{eq9})
and integrating over the momentum,  is given by
\be
P_n=\frac{1}{24\pi^2}\bigg[k_{F}^{n}\mu_n^\star\left(2 k_{F}^{n}-3m^{\star 2}\right)+3m^{\star 4}\log\frac {k_F^n+\mu_n^\star}{m^\star}\bigg]
\ee
The pressure due to the protons, on the other hand, is given by
\be
P_p=P_{\rm field}+P_{med}
\ee
The magnetic field contribution to the pressure $P_{\rm field}$, from the Dirac sea is given by, from Eq.(\ref{omgf})
\bea
&& P_{\rm field}=
\frac{|eB|^2}{(2\pi)^2}\bigg[\zeta'(-1,x)\nonumber\\
&&-\frac{1}{2}(x_p^2-x_p)\log(x_p)+\frac{x_p^2}{4}+\frac{1}{12}\log x_p\bigg].
\label{pref}
\eea

The medium contribution to the pressure from the protons, on the other hand, is given by, 
\bea
&&P_{med}=- \sum_{n=0}^{n_{max}} \frac{{\alpha}_n |eB|}{4\pi^2} 
\Big[ \mu_p^\star k_{F,n}^p\nonumber\\
& - &\Big(m^{\star 2} 
+ 2n|eB| B \Big) \log \left(\frac{k_{F,n}^p + \mu_p^*}{m^{\star 2} + 2n|eB |}\right) \Big]
\label{pmed}
\eea

    In order to account for NS matter, one needs to incorporate the charge neutrality and 
beta equilibrium conditions as well. The charge neutrality conditions are as follows
\begin{eqnarray} 
\sum_{B} Q_B~\rho_B + \sum_{l} Q_l~\rho_l = 0 
\end{eqnarray}
where, the suffix B is summed over nucleons ($n,p$) while suffix $l$ denotes sum over 
all leptonic states ($e,\mu$). $Q_B$ and $Q_l$ are the electrical charges of baryons and the leptons, respectively. 
The $\rho_B$ ($n,p$) and $\rho_l$ ($e,\mu$) are the total baryons and leptons density, respectively. 
Thus, at a given baryon number density $\rho_B =(\rho_n + \rho_p)$, the charge neutrality
condition is given by, $\rho_p=\rho_e+\rho_\mu$. 
The beta equilibrium condition lead to the chemical potentials of the proton, neutron , electron and muons 
given as
$\mu_n=\mu_p+\mu_e$ and $\mu_e=\mu_\mu$
while the number densities of neutron and proton are given in Eq.(\ref{rhopn}), the 
lepton number density $\rho_l$ (electron and muons) is given by
\be
\rho_l=\sum_n^{n_{max}}\alpha_n\frac{|eB|}{2\pi^2}\sqrt{\mu_E^2-2n|eB|}
\ee
where, we have neglected the mass of electrons and in the sum above, the maximum number of Landau levels
is given as $n_{max}={\rm Int}[\mu_E^2/(2|eB)]$.

Let us note that the EOS that we have derived given in eq.(\ref{eosprmag}) corresponds 
to the thermodynamic pressure i.e. negative of the thermodynamic potential. However, in the presence 
of magnetic field the hydrodynamic pressure  can be highly anisotropic when there is significant 
magnetization \cite{Chatterjee2011,Ferrer2010,Huang2010,Ventura1977} of the matter. The pressure in the direction of the 
field $P_{\parallel}$ is the thermodynamic pressure as given in Eq.(\ref{eosprmag}). 
On the other hand, the pressure $P_\perp$ in the transverse direction of the magnetic field
is given by, with $P_0$ as defined in Eq.(\ref{eosprmag}),
\be
\label{eq37}
P_\perp=P_0-{\cal M}B+\frac{1}{2}B^2
\ee
where, ${\cal M}=-\partial\Omega/\partial B$ is the magnetization of the
system. 
Using Eq.(\ref{pref}) and Eq.(\ref{pmed}), the total magnetization can be written as 
${\cal M}={\cal M}_{med}+{\cal M}_{\rm field}$ where,
the magnetization of the medium is given as
\bea
&&{\cal M}_{med}=\frac{\partial P_{med}}{\partial B}=\sum_{n=0}^{n_{max}} 
\frac{{\alpha}_n |e|}{4\pi^2} \nonumber\\
&&\Big[ \mu_p^\star k_{F,n}^p\nonumber\\
&& - \Big(m^{\star 2} 
+ 4n|eB| B \Big) \log \left(\frac{k_{F,n}^p + \mu_p^*}{m^{\star 2} + 2n|eB |}\right) \Big].
\label{mmed}
\eea
On the other hand, the magnetization of the Dirac sea is given as
\bea
&&{\cal M}_{\rm field}=\frac{\partial P_{\rm field}}{\partial B}\nonumber\\
&&=\frac{e^2B}{\pi^2}\bigg[\frac{1}{12}\log x_p-\frac{1}{24}+\frac{x^3}{2}I_2\bigg]
\label{magfield}
\eea
where, we have used the expression for $\zeta'(-1,x)$ given in Eq.(\ref{zetader})
and defined the quantity $I_2$ as the integral\cite{Chatterjee2011}
\be
I_2=(2\pi)
\int_0^{\infty}\frac{2\tan^{-1}y +y\log(1+y^2)}{(\exp(2\pi xy)-1)(1-\exp(-2\pi x y))}y dy
\ee

To obtain the gross structural properties of the neutron star with the equation of state as calculated above,
we also introduce a density dependent magnetic field  \cite{Ghosh2007},
\bea
\label{eq41}
B=B_{\rm surf} + B_0 \left[ 1 - e^{- \beta (\rho_B/\rho_0)^\gamma}  \right]  
\eea
where, $\rho_0=0.14fm^{-3}$ is the nuclear saturation density,
$B_{\rm surf}=10^{15}$ Gauss is the magnetic field on the surface, $B_0$ is the maximum magnetic field at the core.
The parameters $\beta$ and $\gamma$ are chosen 0.003 and 3, respectively \cite{Chu2015}. 
such that the field increases somewhat mildly with density to its core value but still describes correctly the surface,
namely with a zero pressure.

\section{Neutron Star Structure and tidal deformability}
               The equations for the structure of a relativistic spherical and static star composed 
of a perfect fluid were derived from Einstein's equations by Tolman, Oppenheimer and Volkoff, 
known as Tolman-Oppenheimer-Volkoff equations (TOV), which are \cite{Weinberg}
\bea
\label{tov1}
\frac{dP}{dr}=-\frac{G}{r}\frac{\left[\varepsilon+P\right]\left[M+4\pi r^3 P\right ]}{(r-2 GM)},
\eea
\bea
\label{tov2}
\frac{dM}{dr}= 4\pi r^2 \varepsilon,
\eea
\noindent
        with $G$ as the gravitational constant and $M(r)$ as the enclosed gravitational mass. 
For the specified EOS, these equations can be integrated from the origin as an initial 
value problem for a given choice of central energy density, $(\varepsilon_c)$.
The value of $r~(=R)$, where the pressure vanishes defines the surface of the star.
We solve these equations to study the structural properties of a static neutron star using EOS as 
derived and given in Eq.(\ref{eosenmag}) and Eq.(\ref{eosprmag}) for the magnetized, charge neutral dense nuclear
matter.

While simultaneous measurements of mass and radius neutron stars have the potential to
constrain equation of state for the neutron star matter, such measurements are also
plagued with uncertainties and model dependence on the radiation mechanisms at the neutron 
star surface as well as interstellar absorption. On the other hand, observation of inspiralling 
binary neutron stars with the gravitational wave detection GW170817,
could provide significant information about the structure of the neutron stars. 
The tidal distortion of the neutron stars in a binary system links the equation 
of state to the gravitational wave emission during the inspiral \cite{Malik2018,Gomes2018}. 
In the following we shall estimate this parameter for the equation of state
for the magnetized nuclear matter.

The tidal deformity parameter $\lambda$ relates the induced quadrapole moment $Q_{ij}$ of a neutron star due to the strong 
tidal gravitational field ${\cal E}_{ij}$ of the companion star. This qudrupole deformation in leading order
in perturbation is given as \cite{Hinderer2008}
\be
Q_{ij}=-\lambda{\cal E}_{ij}.
\ee
The parameter $\lambda$ is related to the $l=2$, the tidal Love number as $k_2=\frac{3}{2}\lambda R^{-5}$, R being the 
radius of the neutron star. One can estimate $k_2$ perturbatively by estimating the deformation $h_{\alpha\beta}$ of the
metric  from the spherical metric. We consider here, the leading order static perturbation and axisymmetric
perturbation.
The deformation of the metric in Regge-Wheeler gauge can be written as \cite{Hinderer2008}
\bea
&&h_{\alpha\beta}=\nonumber\\
&&diag\Bigg[e^{-2\Phi(r)}H_0,e^{2\Lambda(r)}H_2,r^2K(r), \nonumber\\
&&r^2\sin^2\theta K(r)\Bigg]
Y_{20}(\theta,\phi)
\eea
where, $H_0(r)$, $H_2(r)$ and $K(r)$ are the perturbed metric functions. It turns out that $H_2(r)=H_0(r)\equiv H(r)$
using Einstein's equation $\delta G_\alpha^\beta=\delta T_\alpha^\beta$ while $k'(r)=2H(r)\Phi '(r)$.
The logarithmic derivative of the deformation function $H(r)$ i.e. $y(r)=r\frac{H_0^\prime(r)}{H_0(r)}$ satisfies the
first order equation \cite{Damour2009}
\bea
\label{eq46}
r \frac{d y(r)}{dr} + {y(r)}^2 + y(r) F(r) + r^2 Q(r) = 0,
\eea
with
\bea
F(r) = \frac{r-4 \pi r^3 \left( \epsilon(r) - p(r)\right) }{r-2
M(r)},
\eea
\bea
\nonumber Q(r) &=& \frac{4 \pi r \left(5 \epsilon(r) +9 p(r) +
\frac{\epsilon(r) + p(r)}{\partial p(r)/\partial
\epsilon(r)} - \frac{6}{4 \pi r^2}\right)}{r-2M(r)} \\
&-&  4\left[\frac{M(r) + 4 \pi r^3
p(r)}{r^2\left(1-2M(r)/r\right)}\right]^2 \ .
\eea
To calculate the tidal deformation,
the equation for the metric perturbation, Eq.(\ref{eq46}) can be integrated together with the TOV 
Eqs.(\ref{tov1},\ref{tov2}) for a given EOS radially outwards, with the boundary conditions $y(0) = 2$, 
$p(0)\!=\!p_{c}$ and $M(0)\!=\!0$, Where $y(0)$, $p_c$ and $M(0)$ are the 
the  pressure and the mass density at the center of the NS, respectively.

The tidal Love number $k_2$ is related to $y_R=y(R)$ through
\bea
\label{eq49}
&& k_2 = \frac{8C^5}{5}\left(1-2C\right)^2
\left[2+2C\left(y_R-1\right)-y_R\right]\times \\
&&\bigg\{2C\left(6-3 y_R+3 C(5y_R-8)\right)\nn \\
&&+4C^3\left[13-11y_R+C(3 y_R-2)+2
C^2(1+y_R)\right] \nn \\
&& ~ ~
+3(1-2C)^2\left[2-y_R+2C(y_R-1)\right]\log\left(1-2C\right)\bigg\}^{-1},\nn
\eea
where $C$ $(\equiv M/R)$ is the compactness parameter of the star of
mass $M$.
The dimensionless tidal deformability $\Lambda$ is defined as \cite{Flanagan2008,Hinderer2008,Hinderer2010,Damour2012}, 
 \bea
 \label{eq50}
 \Lambda=\frac{2}{3} k_2 (R/M)^5,
 \eea

The observable signature of relativistic tidal deformation will have an effect on the phase evolution
of the gravitational wave spectrum from the inspiral  binary NS system. This signal will have cumulative effects
of the tidal deformation arising from both the stars. Therefore, one can combine the tidal deformabilities and
define a  dimensionless tidal deformability $\Lambda$ taking a weighted average as \cite{favata2014}
\be
\tilde \Lambda=\frac{16}{13}\left[\frac{(M_1+12 M_2)M_1^4\Lambda_1+(M_2+12 M_1)M_1^4\Lambda_2}{
(M_1+M_2)^5}\right]
\ee
In the above, $\Lambda_1$,$\Lambda_2$ are the individual tidal deformabilities corresponding to
the two components of neutron star binary with masses $M_1$ and $M_2$ respectively.

\section{Results and Discussion} \label{results}  
                As mentioned earlier, we only incorporate the $\rho-\sigma$ cross coupling 
                parameterization of the energy density functional (EDF) as 
                given in Eqs. (\ref{eq8},\ref{eq9}) in the present effective chiral model 
                given in Eq. \ref{eq1}. With this cross coupling term the EDF is able to 
                satisfy the nuclear matter properties, specifically the density dependence 
                of symmetry energy with the available empirical estimates as given in Table 
                \ref{tab1} as well as it also satisfies the present constraints on 
                neutron star properties such as NS maximum mass and radius \cite{Malik2017}. 
                These calculations were performed without the effect of magnetic field. 
                In the following, we show the effects of strong magnetic field on this EDF. 
                In particular, we investigate the effective mass of nucleons, relative
                particle population of charged and uncharged particle 
                in $\beta$-equilibrated nuclear matter. 

\begin{figure}[!ht]
\includegraphics[width=85mm]{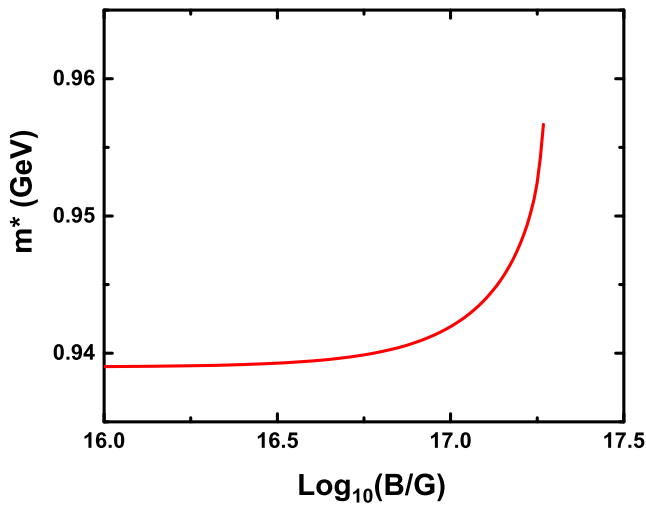}
\caption{\label{fig1}(Color online) The vacuum nucleon mass as a function of magnetic field.}
\end{figure}                            
                 Let us first discuss the vacuum nucleon mass i.e. the nucleon mass at zero baryon density 
                 as a function of magnetic field. For $\mu=0$, the Eqs. (\ref{eq3}-\ref{eq5}) are solved trivially 
                 with $\rho_0^3=0$ and $\omega_0=0$. The remaining Eq. (\ref{eq4}) is to be solve for 
                 the effective mass ($Y=m^*/m$). For zero magnetic field there is a unique solution $Y=1$ i.e. 
                 $m^*=m$. However, because of Dirac sea response to magnetic field, the scalar density for 
                 proton has a non vanishing contributions $\rho_s^{\it field}$ as shown in Eqs. (\ref{eq19},\ref{eq24}).
                 The numerical solutions obtained by solving Eq. (\ref{eq4}) at non zero magnetic field
                 for the effective nucleon mass is shown in Fig. \ref{fig1}. The present model being an effective
                 model for nucleon matter is not expected to be profound results for vacuum of strong interactions. 
                 None the less, we see a magnetic catalysis for nucleon mass similar to magnetic catalysis of 
                 chiral symmetry breaking for quarks \cite{Chatterjee2011}. For very large magnetic 
                field (larger than $eB\sim10m_\pi^2$) 
                 we do not get any solutions for Eq. (\ref{eq4}) for the effective nucleon mass. This can be an 
                 artifact of the mean field approximations that we use here. Going beyond the mean field with a magnetic 
                 field dependent meson masses can possibly cure this \cite{Haber2014}. 
                 In what follows, however, we shall neglect such an effect on the meson masses and continue with the 
                 mean field approximation. We note however that there is no such limitations at finite baryon density. 
                 
\begin{figure}[!ht]
\includegraphics[width=85mm]{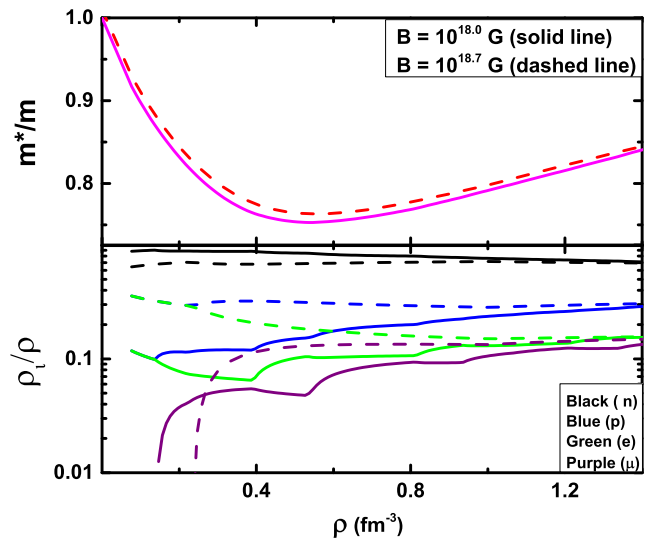}
\caption{\label{fig2}(Color online) Variation of nucleon effective mass (upper panel) and the relative 
particle population (lower panel) for $\beta-$ equilibrated nucleon matter with total baryon density 
for different strengths of magnetic field ($B$).}
\end{figure}                 
                In Fig. \ref{fig2}, in the upper panel, we show the variation of the nucleon mass for constant magnetic field. 
                The effect of magnetic field become significant only when the field strength exceeds about $10^{17}$ Gauss. 
                Due to magnetic catalysis the effective mass in presence of magnetic field is larger than the 
                mass without the magnetic field. On the lower panel, we display the relative populations of the charge particles.
                The populations of the charged particles are influenced both by magnetic field and the charge neutrality 
                condition. The proton fraction remains sufficiently small till $2.5 \rho_0$. As the magnetic field is increased
                protons contribute at still higher density as their masses become heavier with magnetic field.  
                
\begin{figure}[!ht]
\includegraphics[width=85mm]{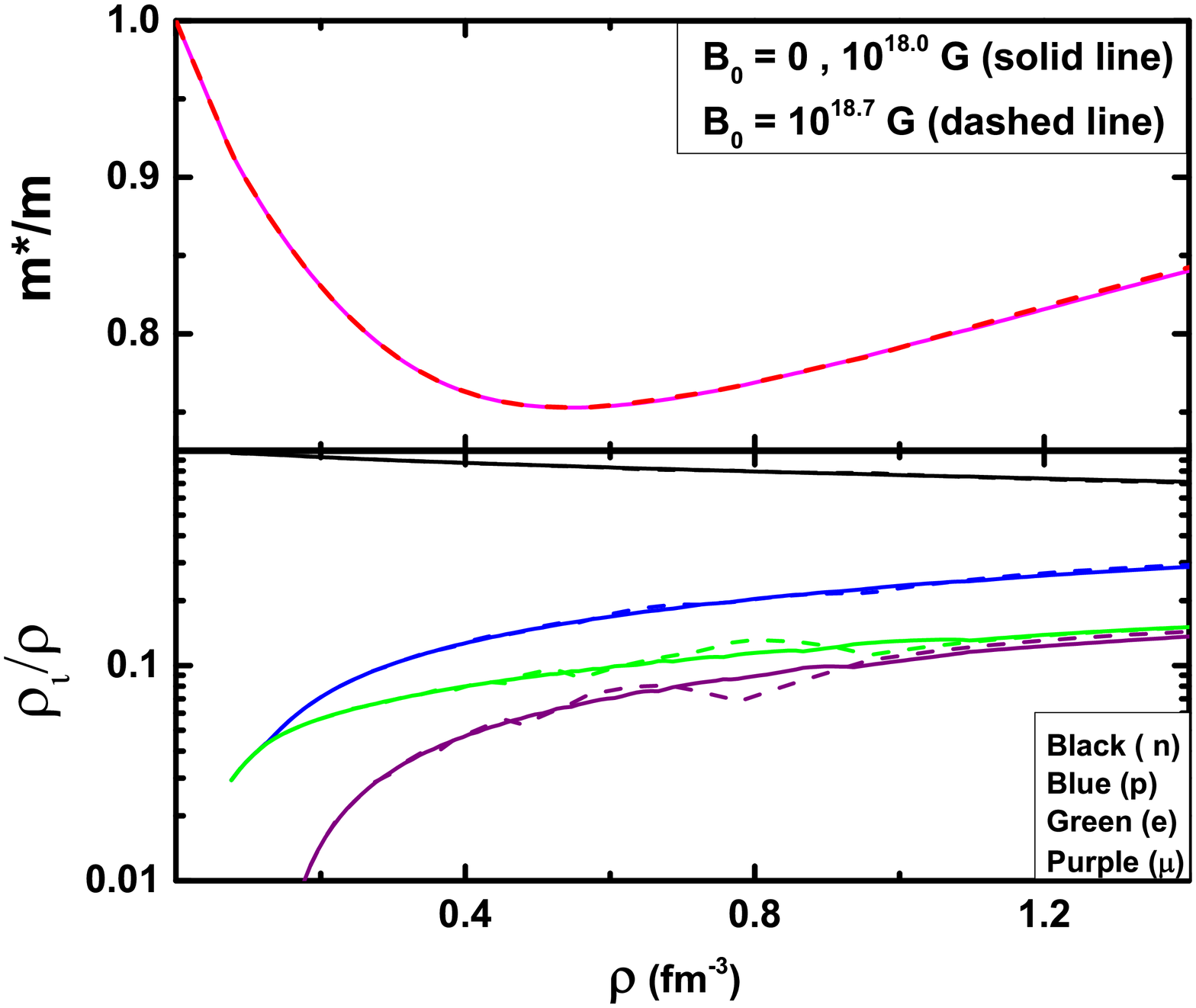}
\caption{\label{fig3}(Color online) Variation of nucleon effective mass (upper panel) and the relative 
particle population (lower panel) for $\beta-$ equilibrated nucleon matter with total baryon density 
for density dependent magnetic field given in Eq. (\ref{eq41}). The solid lines correspond to the  
central value of magnetic field $B_0=10^{18}$ Gauss where as the dashed lines represent 
$B_0=10^{18.5}$ Gauss.}
\end{figure}                
                Next in Fig. \ref{fig3} we show the same variables as in Fig. \ref{fig2} but with a 
                density dependent magnetic field as given in Eq. \ref{eq41}. 
                As may be seen in the figure, the proton fraction remains sufficiently small
                up to 3-4 times nuclear matter density and the contribution of the neutrons 
                to the pressure remain dominant. As compared to Fig. \ref{fig2} the effect of 
                magnetic field in this case is milder because the field decreases as the density 
                decreases. The strength of the magnetic field to induce 
                significant changes can be estimated in a straight forward manner. The contributions 
                from the protons become significant when the lowest Landau level ($n=0$) is occupied.  
                One can estimate this to happen when $2eB > \sqrt{(\mu_p^{\star 2}-m^{\star 2})}$. 
                Equivalently, this correspond to a magnetic field $eB> 3.2\times 10^{19} (\rho_p/\rho_0)^{2/3}$ 
                Gauss where, $\rho_0\sim 0.16 fm^{-3}$. This is the reason why the effect of magnetic 
                field is not seen for magnetic fields up to $B_0\sim 10^{18}$ Gauss. 
                Moreover, with the density dependence as in Eq.(\ref{eq41}), the effect is seen 
                only at high density. It can be noted that both effective mass and hence the population 
                of different particles do not show any changes for the magnetic field intensity 
                at NS core within $B_0 =$ $0-10^{18}$ Gauss whereas, there is a noticeable increase in 
                both nucleon effective mass and proton population when the $B_0$ is more than $10^{18}$ Gauss.
                \cite{Broderick2000}. Similar effect can be seen in the relative 
                particle population when $B_0 > 10^{18}$ Gauss. There is an increment in the 
                charge particle concentration for the charged species, such as $e^-, p^+$ and $\mu^-$ at 
                $\approx$ $\rho = 3 \rho_0$, as a result of which the neutron concentration drops 
                at higher densities, in comparison to the no field case or NS core field intensity 
                till $B_0 \leq 10^{18}$ Gauss.
\begin{figure} [!ht]
\includegraphics[width=85mm]{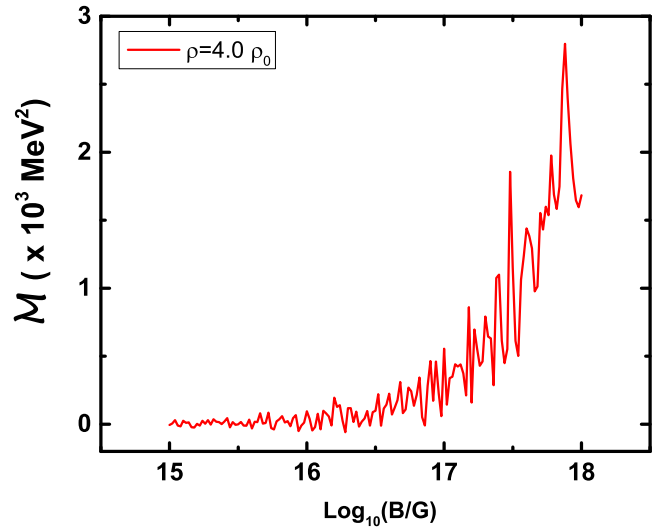}
\caption{\label{fig4} (Color online) Variation of total magnetization with magnetic field. 
We have taken here $\rho=4.0\rho_0$.} 
\end{figure} 
      The net magnetization in the matter, as given in Eq. (\ref{mmed}) {and Eq. \ref{magfield}}, 
      is shown in Fig. \ref{fig4}. It shows that the net magnetization 
      is almost negligible till $B=10^{16}$ Gauss. From $B\geq10^{17}$ Gauss onwards it starts 
      increasing and becomes highly oscillatory. The oscillation of the magnetization happens 
      as an outcome of the well-known de Haas-van Alphen effect 
      \cite{Ebert1994,Shoenberg1984,Chatterjee2011} 
      in which the charged particles, due to Landau quantization, exist only in orbitally quantized 
      states in a magnetic field and as the number of occupied Landau levels changes with 
      the magnetic field, the magnetization becomes oscillatory. The oscillatory behavior is 
      more pronounced with the increasing strength of the magnetic field.  The irregularity 
      in the oscillation is due to the medium dependence of the nucleon mass which itself depends 
      on the magnetic field. This behavior is also consistent with the findings of 
      \cite{Broderick2000,Chatterjee2011}.
\begin{figure} [!ht]
\includegraphics[width=80mm]{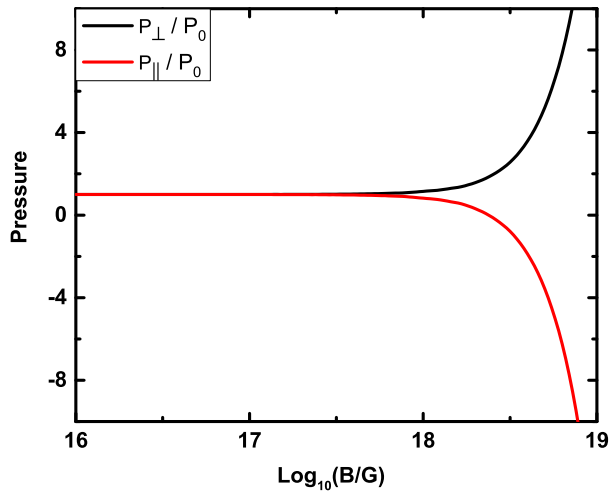}
\caption{\label{fig5} (Color online) Variation of pressure (scaled with no field value $P_0$) in parallel and 
perpendicular directions of magnetic field with respect to magnetic field at baryon 
density 4.0$\rho_0$.} 
\end{figure} 
               We also show the variation of pressure (scaled with the value $P_0$, pressure in absence of magnetic field) 
               in parallel and perpendicular directions of magnetic field with respect to magnetic field intensity 
               at baryon density 4.0$\rho_0$ in Fig. \ref{fig5}. The parallel component of the pressure as given in 
               Eq. (\ref{eq13}) decreases with magnetic field and even becomes negative as the magnetic field is 
               increased. On the other hand the perpendicular component as given in Eq. (\ref{eq37}) monotonically increases 
               with magnetic field. It turns out that at this density $(\rho=4.0 \rho_0)$, the magnetization conurbations ${\cal M} B$ 
               is two orders of magnitude smaller compared to the matter contribution. Therefore the oscillatory behavior of 
               magnetization as seen in Fig. \ref{fig4} is not reflected in the transverse pressure in Fig. \ref{fig5}.                
               For $\rho=4.0\rho_0$, the parallel pressure starts becoming beyond $e B=10^{18.4}$ Gauss.  

\begin{figure} [!ht]
\includegraphics[width=85mm]{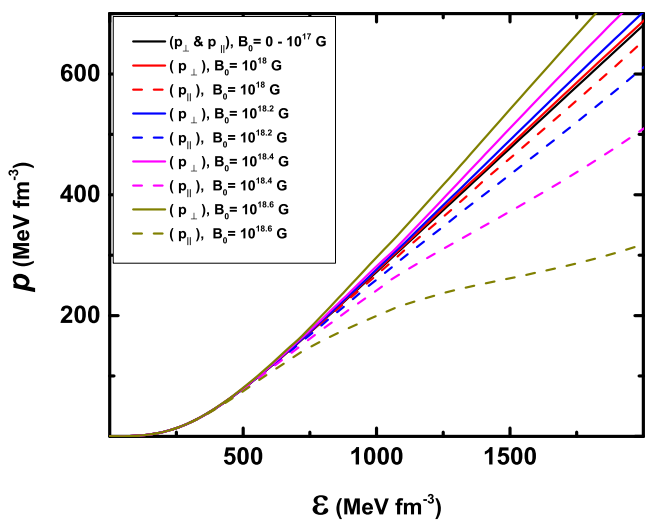}
\caption{\label{fig6} (Color online) Variation of pressure vs energy density for different values of central 
magnetic field ($B_0$). The solid lines correspond to $P_\perp$ while the dashed lines correspond 
to $p_\parallel$.} 
\end{figure}     
            Next In Fig. \ref{fig6}, we show the EOS of magnetized charged neutral matter. Here we have taken 
            density dependent magnetic field given in Eq. (\ref{eq41}). Similar to Fig. \ref{fig2} for 
            the effective masses, the pressure does not show any change as the field strength is increased 
            up to $B_0=10^{17}$ Gauss, the magnetic field at the core of the NS. The effects of the magnetic field 
            become noticeable for $B_0$ beyond $10^{18}$ Gauss. In the Figure, the solid and dashed lines correspond
            to the pressure in the perpendicular and the parallel direction to the magnetic field. Beyond
            $10^{18}$ Gauss the difference between these two pressures increases rapidly. As is obvious from the figure
            the perpendicular component become stiffer while the the parallel component become softer. For $B_0$ beyond 
            $10^{19}$ Gauss the parallel component become negative. Later we shall be using TOV Eqs. (\ref{tov1},\ref{tov2})
            to solve for mass and radius of magnetized NS, we keep magnetic field strength $B_0$ up to $10^{18}$ Gauss
            so that the anisotropy of pressure is not two large and the spherically symmetric TOV equations can be applicable. 

\begin{figure} [!ht]
\includegraphics[width=85mm]{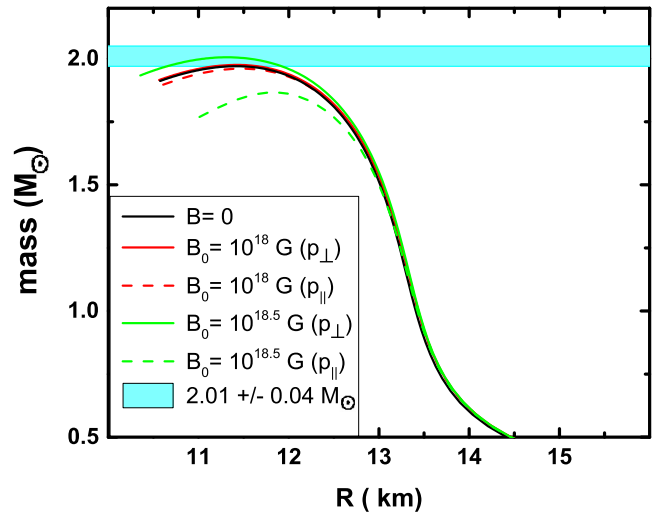}
\caption{\label{fig7} (Color online) The mass and radius relationship for different central magnetic field $B_0$. 
The dashed line correspond to taking $P_\parallel$ while the solid line correspond to $P_\perp$.
         Maximum mass limits imposed from recent observation of high mass star PSR J0348 - 0432 
         ($M=(2.01 \pm 0.04) M_{\odot}$) \cite{Antoniadis2013} (cyan band) is also indicated.}
\end{figure}
             
                       We then proceed to calculate the mass-radius relationship with the magnetized charge neutral 
             neutron star matter, which is shown in Fig. \ref{fig7}. The TOV equations are solved by taking the EOS 
             with both parallel and perpendicular pressure. The solid line in the figure correspond to taking 
             the $P_\perp$ for pressure in TOV equations. The dashed line correspond to taking the pressure as $p_\parallel$. 
             We find that there is no appreciable change in the gravitational mass and radius of the 
             neutron star when magnetic effects are incorporated for field straight up to $B_0=10^{17}$ Gauss,
             compared to the case where magnetic field is absent. 
             Even with the increase in central magnetic field up to $10^{18}$ Gauss 
             there is hardly any change in the gross structural properties of 
             the neutron star although the particle concentration 
             in the neutron star matter changes as seen in the lower panel of Fig. \ref{fig3}. 
             Increasing the value of $B_0$ further results in making anisotropy between the two 
             pressure larger with even the parallel component of the pressure becoming negative
             leading to mechanical instability. For central value of the magnetic field $B_0$
             up to $10^{18}$ Gauss the resulting maximum mass of the neutron star satisfy 
             the maximum mass constraint ($M=2.01 \pm 0.04 M_{\odot}$) \cite{Antoniadis2013}
             in both $P_\perp$ and $P_\parallel$ as shown in figure.
             We have also shown the results for $B_0=10^{18.5}$ Gauss. The maximum mass corresponding to
             $P_\perp$ and $P_\parallel$ are $2.01 M_{\odot}$ and $1.87 M_{\odot}$ with asymmetry $\delta$
             in the masses ($\delta=(M_\perp-M_\parallel)/(M_\perp+M_\parallel)$) to be about $3.6\%$.
             This only means that the pressure anisotropy is rather small leading to tiny mass asymmetry and hence 
             gives a post-facto justification of using the isotropic TOV equations to calculate the gross
             structural properties of NS approximately.  
             The corresponding radii ($R_{1.4}$) of the canonical mass neutron star ($M=1.4 M_{\odot}$) 
             that we obtain in the present model also agrees well the empirical estimates given by \cite{Lattimer2013}. 
             The overall results obtained from the solutions of TOV equations are tabulated in Table \ref{tab2}.
             For the central magnetic field of $10^{18}$ Gauss by taking $P_\parallel$ in TOV equations we get a lower maximum 
             mass which is probably expected as the corresponding EOS becomes softer for $P_\parallel$.
             
\begin{table}[ht!]
\caption{\label{tab2}Neutron star properties such as the mass ($M$), radius ($R$) and canonical radius ($R_{1.4}$) 
for the model under consideration in the perpendicular direction ($\perp$) and parallel direction ($\parallel$)
of magnetic field for different values of $B_0$.}
\begin{tabular}{cccccccccccc}
\hline
\hline
\multicolumn{1}{c}{$B_0$}&
\multicolumn{1}{c}{$M$} &
\multicolumn{1}{c}{$R$} & 
\multicolumn{1}{c}{${R_{1.4}}$} \\
%
\multicolumn{1}{c}{(Gauss)}&
\multicolumn{1}{c}{($M_{\odot}$)} &
\multicolumn{1}{c}{($\rm km$)} &
\multicolumn{1}{c}{($\rm km$)} &
\multicolumn{1}{c}{} &
\multicolumn{1}{c}{} \\
\hline 
$0$  &1.97  &11.42 & 13.11 \\
\hline
$10^{15}-10^{17}$ ($\perp$) &1.97  &11.43 & 13.13 \\
~~~~~~($\parallel$)         &1.97  &11.43 & 13.13 \\
\hline
$10^{18}$ ($\perp$) &1.97  &11.43 & 13.14 \\
~~~~~~($\parallel$)         &1.96  &11.47 & 13.13 \\
\hline
$10^{18.5}$ ($\perp$) &2.01  &11.30 & 13.15 \\
~~~~~~($\parallel$)   &1.87  &11.83 & 13.12 \\
\hline
\end{tabular}
\protect
\end{table}
            
\begin{figure} [!ht]
\includegraphics[width=85mm]{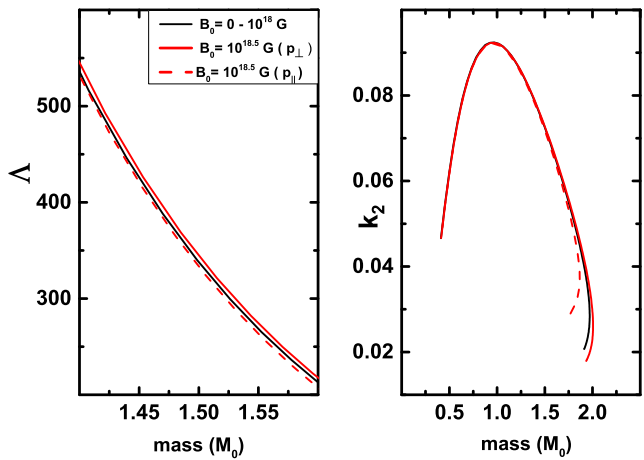}
\caption{\label{fig8} (Color online) (left panel) The tidal deformability ($\Lambda$)  and (right panel) the second Love 
         number ($k_2$) with respect to NS mass for different values of core magnetic field $B_0$.} 
\end{figure}
           
                  As discussed earlier, lately the data from gravitational wave detection from observation of 
           inspiralling binary neutron stars GW170817 \cite{GW170817} could possibly constrain not only the 
           properties of neutron stars but could also put constraints on the EOS. In Fig. \ref{fig8} 
           we plot the dimensionless tidal deformability $\Lambda$ (left panel) as given in Eq. \ref{eq50} 
           and the Love number $k_2$ (right panel) as given in Eq. \ref{eq49} as a function of NS 
           mass for the EOS with different values of central magnetic field ($B_0$). Let us note that the Love 
           number $k_2$ not only depends upon the compactness parameter $C\equiv M/R$ but also on $y(R)$, the 
           value of logarithmic derivative of the deformation function which depends upon internal structure
           of the NS as in Eq. \ref{eq46}. The value of $k_2$ has a peak around $1.0 M_{\odot}$ while it is 
           rather low at higher and lower masses as it is seen in right panel of the figure indicating 
           that the quadrapole deformation is maximum for intermediate mass ranges for a given EOS.  
           The obtained value of $\Lambda$ for $1.4~{\rm M}_\odot$ NS is $520$, 
           in the present model without inclusion of the magnetic field. 
           The $\Lambda$ and $k_2$ are 
           also calculated for both the EOSs with pressure perpendicular and parallel direction to the 
           the magnetic field. There is almost no change in their values for 
           $B_0=0-10^{18}$ G with EOS for both the cases. 
           However, 
           for $B_0>10^{18}$ the $\Lambda$ and $k_2$ values both increases in the perpendicular direction and the effect is 
           opposite for the parallel case. However the change is very small and for $k_2$ the effect is only seen for larger 
           NS masses. Thus the magnetized NS still satisfies the constraint on the tidal deformability parameter 
           $\Lambda$ bound from gravitational wave data GW170817. The corresponding love number ($k_2$) also 
           lies within the acceptable range \cite{Hinderer2008}. 
           The agreement of both these parameters validates the properties of neutron star obtained within the 
           model with the inclusion of magnetic field.

\begin{figure} [!ht]
\includegraphics[width=85mm]{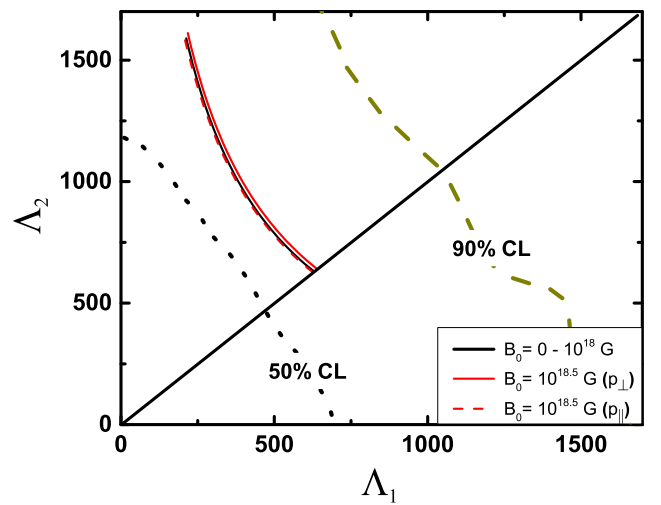}
\caption{\label{fig9} (Color online) Tidal deformabilities associated with individual components of the binary 
of GW170817 using the magnetized EOS. The $50\%$ and $90\%$ confidence limits for this event 
are also indicated.} 
\end{figure} 
             Finally, in Fig. \ref{fig9} we plot the tidal deformability parameters $\Lambda_1$
             and $\Lambda_2$ which are linked to the neutron star binary companion having a high 
             mass $M_1$ and a low mass $M_2$ associated with GW170817 event. We have plotted them 
             for the EOS considered here with magnetic field. The curves are obtained by 
             varying the high mass ($M_1$) independently in the range $1.365 < M/M_{\odot} < 1.60$ 
             obtained for GW170817 whereas the low mass ($M_2$) is determined by keeping the 
             chirp mass ($M_{chirp}= (M_1 M_2)^{(3/5)}/(M_1+M_2)^{5}$) fixed at the observed 
             value $1.188 M_{\odot}$. The long dashed dark yellow line signifies the $90\%$ 
             probability contour found from this event. The black doted line on the other hand 
             signifies the $50\%$ probability contour. As may be observed from the figure 
             the EOS obtained from the present chiral model for nucleon matter lies well within 
             the two limits with or without the magnetic field.

\section{Conclusions}
          Let us summarize the salient feature of the present investigation. 
          We have looked into the different effects of magnetic field on a 
          nuclear matter EOS within the ambit of a chiral model which is a generalized
          sigma model couple to nucleons. Apart from sigma and pions the model incorporate
          $\rho$ and $\omega$ mesons with higher order mesonic fields. The model with cross 
          coupling between isovector and scalar field successfully described symmetry energy 
          parameters \cite{Malik2017}. 
          
          In the present calculation of incorporating the effects of magnetic field, we have included
          the effect of the field on the Dirac sea of protons. This effects the vacuum mass of the nucleon 
          at zero baryon density and temperature. The mass of the nucleons are seen to be 
          increasing with magnetic field showing the magnetic catalysis effects. For small field 
          this increase in mass is seen to be quadratically dependent on the magnetic field which we take to be 
          homogeneous. 
          
          Next we calculated the effective mass of the nucleons in the medium with non zero densities
          in the presence of constant magnetic field. For field straight up to $10^{17}$ Gauss the EOS 
          does not change as compare to the case for vanishing magnetic field. Due to the directional 
          dependence of the magnetic field the pressure is no longer isotropic. The EOS for the pressure 
          parallel to the magnetic field ($P_\parallel$) become softer while the same perpendicular to the 
          magnetic field ($P_\perp$) becomes stiffer. For small anisotropy in pressure we have used TOV 
          equations to obtain gross structural properties of the NS. Within this approximations the correction
          due to the magnetic field remains small for the masses and radii of a NS. The masses of the NS 
          and their radii appeared to be within the corresponding acceptance limits.
          
          We have next estimated the tidal deformability, the Love number for NS with and without 
          magnetic field. For the strength of magnetic field considered here ($B_0=10^{18.5}$ Gauss) the 
          tidal deformability parameters do not differ much from the zero magnetic field cases and 
          lie within the acceptable range for $\Lambda$ from GW170817. We have also calculated the 
          tidal deformabilities in the phase space of $\Lambda_1$ and $\Lambda_2$ associated with
          the two component of the binary related to GW170817. The present EOS with or without magnetic 
          field is consistent with the limits derived from this event. 
          
          We have confined our attention to the EOS based on nucleonic degrees of freedom in the 
          present investigation. It will be interesting to study the effect of magnetic field 
          on the models of NS having a hyperonic core or even a quark core with different exotic
          phases. Some of these investigation are in progress and will be reported elsewhere. 

\begin{acknowledgements}
TM would like to acknowledge kind hospitality provided by Physical Research Laboratory, 
Ahmedabad, where a part of this work was done. TKJ \& TM would like to acknowledge financial 
support from BRNS Project (BRNS/37P/5/2013).
\end{acknowledgements}


\end{document}